\documentclass[conference,letterpaper]{IEEEtran}

\usepackage{dsfont}
\usepackage{amsthm}
\usepackage{amsmath} %
\usepackage{mathtools}
\usepackage{graphicx}
\usepackage{psfrag}
\usepackage{epsf,epsfig}
\usepackage{epstopdf}
\usepackage{subfigure}

\usepackage{caption}
\usepackage{subcaption}

\usepackage{cite,color}
\usepackage{cite}
\usepackage{amsmath,amssymb,amsfonts}
\usepackage{algorithmic}
\usepackage{graphicx}
\usepackage{textcomp}
\usepackage{xcolor}
\usepackage{amsmath}

\usepackage{multirow}

\usepackage{soul}

\definecolor{red}{rgb}{1,0,0}  

\newcommand{\Ncal}{\mathcal{N}}

\newcommand{\Ibf}{{\bf I}}

\newcommand{\Qbf}{{\bf Q}}

\newcommand{\Rbf}{{\bf R}}

\newcommand{\snr }{{\rm snr }}

\usepackage{thmbox}

\begin{document}
\title{Multi-Modal Concurrent Transmission }
\vspace{-.5cm}

%
%
%


\author{
\IEEEauthorblockN{ Majid Nasiri Khormuji\IEEEauthorrefmark{1},
Alberto Giuseppe Perotti\IEEEauthorrefmark{1}, Qin Yi\IEEEauthorrefmark{2} and
Branislav M. Popovic\IEEEauthorrefmark{1}}\\
\vspace{-.5cm}
\IEEEauthorblockA{ \IEEEauthorrefmark{1}\textit{Huawei Technologies Sweden AB, Stockholm, Sweden}
\IEEEauthorrefmark{2}\textit{Huawei Technologies Co., Ltd., Shanghai, China }\\
                           \{majid.nk, alberto.perotti, qinyi4, branislav.popovic\}@huawei.com
                          \vspace{-.3cm}
                           }
}
\maketitle

\begin{abstract}
This paper introduces a novel physical-layer method labelled as  Multi-Modal Concurrent Transmission (MMCT)  for efficient transmission of multiple data streams with different reliability-latency performance requirements. The MMCT arranges data from multiple streams within a same physical-layer transport block wherein stream-specific modulation and coding scheme (MCS) selection is combined with \emph{joint} mapping of modulated codewords to Multiple-Input Multiple-Output spatial layers and frequency resources. Mapping to \emph{spatial-frequency} resources with higher Signal-to-Noise Ratios (SNRs) provides the required performance boost for the more demanding streams.
In tactile internet applications, wherein haptic feedback/actuation and audio-video  streams flow in parallel, the method provides significant SNR and spectral efficiency enhancements compared to conventional 3GPP New Radio (NR) transmission methods.
\end{abstract}


\section{Background, Contributions and Outline}\label{sec:intro}
\vspace{-.1cm}

Tactile internet \cite{Metaverse,Survey_TI,Fettweis}  represents the next evolution in global networking with responsive digital experiences. It facilitates human-to-machine interaction through haptic sensations while simultaneously revolutionizing machine-to-machine interactions. Moreover, it opens the doors for real-time interaction between humans, machines, and their environment. Tactile feedback extends its applications  beyond audiovisual interactions and extends to real-time communication in robotic systems. In the latter case, robots can receive signals that activate and control specific movements. Incorporating tactile information, such as haptic feedback, alongside traditional communication modes, leads to multi-modal transmission. In this paper, a multi-modal transmitter is defined as a transmitter sending  a set of complimentary multiple information streams describing different features (i.e. modes) of the same event.

%

Within immersive multi-modal Virtual Reality (VR) applications, several crucial performance criteria  are defined in \cite{TR_tactile}.
 Notably, the haptic feedback data demands a significantly higher level of reliability, resulting in a lower Block Error Rate (BLER) compared to audiovisual (AV) data. Ensuring the effectiveness of the tactile internet hinges on the reliable reception of this haptic feedback.
%
However, in New Radio (NR) high data rate services (such as enhanced MultiMedia Broadband (eMBB)) and high reliability services (as considered  in  Ultra-Reliable Low Latency Communications (URLLC)) are treated as  two distinct categories. This division poses a challenge in maintaining a seamless interplay between eMBB and URLLC packet flows originating from and addressed to the same application. Consequently, {simultaneously} achieving the required synchronization accuracy and reliability necessitates aggregating data streams at higher layers and transmitting them through a service capable of meeting the most stringent requirements,
as also noted in \cite{Hou}.

In this paper, we present a new multi-modal transmission method that considers the BLER and delay as the main Key Performance Indicators (KPIs) by devising appropriate  resource allocation schemes across space, time, and frequency dimensions that fulfill those requirements. That is, we show how the data of different types should be mapped to  a shared pool of time-frequency-space resources, enabling  \emph{concurrent} transmission while upholding   BLER and delay constraints without compromising the system's spectral efficiency. The analytical and simulation results indicate that it is practically feasible to create two concurrent haptic and video streams, where the haptic stream is significantly more robust to the channel errors while maintaining  almost similar performance for video stream as that in the NR standard.

The paper is organized as follows. Section~\ref{sec:Shortcoming_NR} first discusses the shortcomings of the current NR standard. Section~\ref{sec:Solution} then describes our new Multi-Modal Concurrent Transmission (MMCT). Section~\ref{sec:Peformance_evaluation} discusses both analytical results and  representative Monte-Carlo simulations to showcase the gain of the new method.   Section~\ref{sec:concl} finally concludes the paper.

%

\section{Shortcomings of NR}\label{sec:Shortcoming_NR}
\vspace{-.1cm}


A natural method to apply NR is to multiplex  data from multiple
streams  within each physical layer transport block.
A corresponding shortcoming is that a common Modulation and Coding Scheme (MCS) must be selected to meet the most stringent KPIs among all streams{,} e.g., the lowest BLER which corresponds to the haptic data in this case. This means that a lower MCS should be selected for both types of data to meet their BLER requirements. This therefore reduces the spectral efficiency of the data of less stringent KPI (i.e. video data).   In other words, for a given amount of data and available resources (which results to a given MCS) we need  a higher-SNR link to meet the lower BLER.



Another method is to transmit the data streams independently, with a
	packet scheduler taking care of allocating each packet to a  set of channel resources. The haptic packets arrive almost in every slot for a typical setup, because there may be at most 60 haptic sensors and each follows a Generalized Pareto Distribution  with 100--500 packets/second on average \cite{HW}. In order to guarantee the transmission within delay budget, we need to transmit haptic and video data in the same slot.
%
%
%
%
%
%
The current NR networks are capable of scheduling two Transport Blocks (TBs) with the same {Physical Downlink Shared Channel (}PDSCH{)}. Each TB can have a different MCS.  However, they share the same assigned common time-frequency resource allocations making it unsuitable for transmission of haptic and AV packets with significantly different sizes.

\section{Proposed MMCT Solution}\label{sec:Solution}

Fig.~1 depicts the transmitter functionality encompassing  two data types: haptic feedback data and video data, which are integral data components  of the tactile internet. The block diagram emphasizes key modifications when compared to the NR standard\footnote{To obtain NR transmitter from Fig.~1, we need to remove one MCS, to process both incoming streams with a common MCS and also  to remove the frequency permutation but keeping the layer mapper.}, highlighting adjustments for enabling this scheme.
 Furthermore, we emphasize that the layer mapper, while existing in the NR standard,  has been specifically tailored to suit the MMCT scheme.  
Throughout this discussion, the subscripts  $v$ and $h$ are used to label the corresponding bits, symbols and signals of video and haptic data, respectively. The MMCT scheme is described in the following steps.

\subsection*{\textbf{Step 1: Data-Type Dependent MCS}}\label{sec:Step1}
Each data packet is encoded and modulated independently. That is, the haptic feedback data is processed according to its own MCS and the video data is processed according to its  corresponding MCS.
%
The  modulated symbols for haptic and video  are arranged in segments that are denoted by $\mathbf{x}_{h,i}$ and $\mathbf{x}_{v,j}$, where indices  $i$ and $j$
denote a corresponding set of $n_s \times n_o$ Resource Elements (REs) where $n_s$ is the number of subcarriers in a Resource Block (RB) and $n_o$ is the number of {Orthogonal Frequency-Division Multiplexing (}OFDM{)} symbols in a slot.
The coded and modulated symbols for the haptic (video) over RB $i$ are then given by the following matrix whose entries are enumerated using $i$, $n_{s}$ and $n_{o}$: 
\begin{equation}\label{eq:RB_haptic}
\mathbf{x}_{h,i} =
\begin{pmatrix}
x_{h,((i-1)n_{s}+1,1)} & \cdots & x_{h,((i-1)n_{s}+1,n_{o})} \\
x_{h,((i-1)n_{s}+2,1)}  & \cdots & x_{h,((i-1)n_{s}+2,n_{o})} \\
\vdots  &  \ddots & \vdots  \\
x_{h,(in_{s}-1,1)} &\cdots & x_{h,(in_{s}-1,n_{o})}\\
x_{h,(in_{s},1)} & \cdots & x_{h,(in_{s},n_{o})} \\
\end{pmatrix}.
\vspace{.005cm}
\end{equation}
For a typical NR configuration we can consider that the set of REs are matrices of size 12 by 12; i.e. 12 subcarriers and 12 OFDM symbols as we assume that the two initial OFDM symbols in each slot are used for pilot and/or control information transmission.
The corresponding symbol block for the video data is similarly constructed as in \eqref{eq:RB_haptic}, where the subscript $h$ is changed to $v$.
\setcounter{equation}{2}

\subsection*{\textbf{Step 2: Multi-Data Type Layer Mapper}}\label{sec:Step2}
 Using the set of modulated symbol blocks $\{\mathbf{x}_{h,i}\}$ and $\{\mathbf{x}_{v,j}\}$, we then construct the layer mapper $\mathbb{X}$. Let us denote each spatial layer with index $l$, where the total number of spatial layers is $L$ and the total number of RBs of size $n_{s}\times n_{o}$ is $B$. The layer mapper $\mathbb{X}$ is parameterized by two parameters: $L_h$  and $B_1$.  The parameter $L_h$ ($1\leq L_h\leq L$) denotes the number of highest SNR layers used for the
haptic data transmission  and the parameter $B_1$ denotes the number of RBs used for haptic data over layer $l=L_h$. The remaining $B-B_1$  RBs over layer $l=L_h$ and all the RBs over the layers $L_h+1, L_h+2, \ldots, L$ are used for video data.

The coded modulated symbols are arranged in a 3D array $\mathbb{X}$  in space and frequency, which is given in \eqref{MMCT_Map} on the top of the next page
where the matrix $\mathbb{X} (L_h,B_1) $ has size $n_{s} B\times  n_{o} L$. The haptic data mapping starts with the highest-SNR layer and continues until all symbols are mapped.  Here,  $L_h$ layers are used for  haptic data transmission. The number of RBs used to transmit the haptic data is given by
\begin{align}
N_{RB_{h}}=(L_h-1)B + B_1.
\end{align}
That is, for given assigned resources, the highest SNR-layers are first filled until the haptic data is completely mapped.  Therefore, the total RBs used for the video is  given by
\begin{align}
N_{RB_{v}} =(L-L_h+1)B-B_1.
\end{align}

\begin{figure}
\centering
\vspace{-0.0cm}
	\includegraphics[width=.485\textwidth]{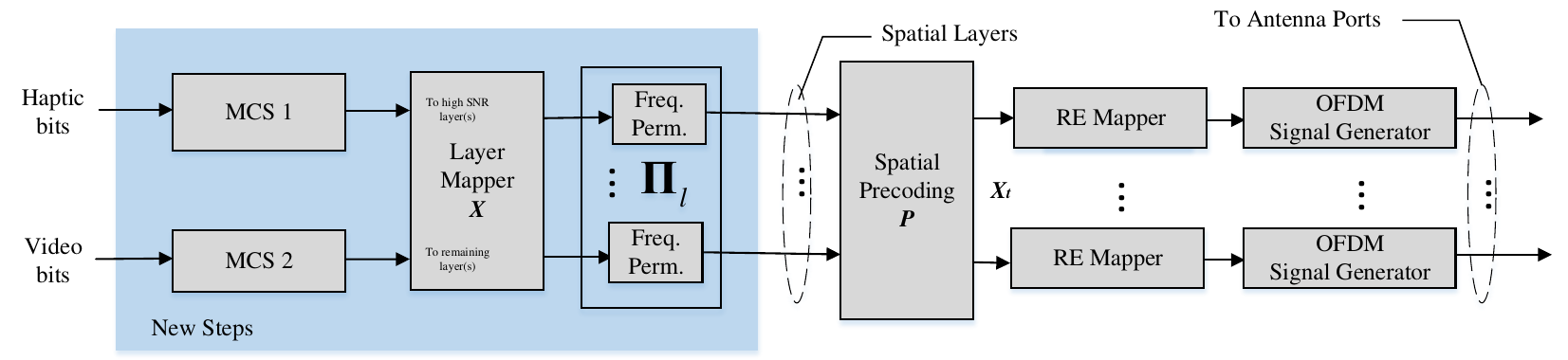}  
\vspace{-0.5cm}
	\caption{The multi-modal concurrent tactile transmitter.}
	\vspace{-0.3cm}
\end{figure}

\newcounter{mytempeqncnt}
\begin{figure*}[t]
\normalsize
\setcounter{equation}{1}
\begin{equation}\label{MMCT_Map}
\mathbb{X} (L_h,B_1)  = \mathcal{M} (\{\mathbf{x}_{h,i}\},\{\mathbf{x}_{v,j}\})=
\begin{pmatrix}
 \mathbf{Layer \ 1 }& \cdots &  \mathbf{Layer \ L_h} &  \mathbf{Layer \ L_h+1} &  \cdots  & \mathbf{Layer \ L}\\
\overbrace{\ \ \mathbf{x}_{h,1} \ \ } &  \cdots & \overbrace{ \ \mathbf{x}_{h,(L_h-1)B+1}   \ }   &\overbrace{  \ \ \mathbf{x}_{v,B-B_1+1}\ \ } &  \cdots & \overbrace{ \mathbf{x}_{v,(L-L_h)B-B_1+1} } \\
\mathbf{x}_{h,2}  & \cdots & \mathbf{x}_{h,(L_h-1)B+2} &   \mathbf{x}_{v,B-B_1+2} &  \cdots & \mathbf{x}_{v,(L-L_h)B-B_1+2}\\
\vdots  & \vdots  & \vdots & \vdots &  \vdots &  \vdots  \\
\mathbf{x}_{h,B_1}&  \cdots & \mathbf{x}_{h,(L_h-1)B+B_1} & \mathbf{x}_{v,B} &  \cdots  & \mathbf{x}_{v,(L-L_h)B} \\
\mathbf{x}_{h,B_1+1}& \cdots & \mathbf{x}_{v,1} & \mathbf{x}_{v,B+1}&  \cdots & \mathbf{x}_{v,(L-L_h)B +1}\\
\vdots  & \vdots  & \vdots & \vdots &  \vdots  &  \vdots \\
\mathbf{x}_{h,B}&  \cdots & \mathbf{x}_{v,B-B_1} & \mathbf{x}_{v,2B-B_1}&  \cdots & \mathbf{x}_{v,(L-L_h)B -B_1}\\
\end{pmatrix}
\end{equation}
\vspace{-.5cm}
\hrulefill
\vspace*{0pt}
\end{figure*}
\setcounter{equation}{4}

\subsection*{\textbf{Step 3: Data-Type Dependent Frequency Permutation}}\label{sec:Step3}
In this step, a layer-specific frequency permutation is done on layer-mapped data symbols $\mathbb{X}$ in \eqref{MMCT_Map}. The purpose of the permutation is to allocate the haptic data symbols to high SNR RBs in each layer to meet the BLER requirement. The permutation is done RB-wise. For layer $l$, the corresponding permutation matrix $\mathbf{\Pi}_l$ is determined by SNR per RB  which can be obtained by channel estimation using Sounding Reference Symbols (SRS) in the Time-Division Duplex (TDD) transmission mode.
One possibility to the design of the permutation over layer $l$ (denoted as $\mathbf{\Pi}_l$) is to restrict the mapping  to depend only on the SNR values of the given layer such that
\begin{align}
\mathbf{\Pi}_l = \mathbb{F}_l \left( \left\{\text{SNR}_{b,l}\right \}_{  1\leq b\leq B } \right),
\end{align}
where $\text{SNR}_{b,l}$ denotes SNR of RB $b$ and layer $l$.

One favorable construction of the permutation matrix $\mathbf{\Pi}_l$ is found using the SNR values for layer $l=L_h$ since this layer is shared between haptic and video data when $1 \leq B_1< B$. The first $B_1$   columns of the permutation matrix $\mathbf{\Pi}_{L_h}$ are formed by selecting the indices of rows of non-zero components $B_1$  of the leftmost columns of $\mathbf{\Pi}_{L_h}$ indicating the RBs with the highest SNRs of the layer. The remaining columns of the matrix $\mathbf{\Pi}_{L_h}$   are chosen in a way to construct a complete permutation matrix. That is, in the layer $L_h$ the permutation matrix is designed so that the haptic data is mapped to $B_1$ RBs having the highest SNRs among all $B$ RBs. For other layers the permutation matrix can be arbitrary, including the identity matrix such that $\mathbf{\Pi}_l=\mathbf{I}$ for $l \neq L_h$, where $\mathbf{I}$ denotes an identity matrix of an appropriate size.

To illustrate an exemplary implementation, consider 8 RBs wherein 3 RBs are allocated for haptic data denoted by $\{\mathbf{a}_h,\mathbf{b}_h,\mathbf{c}_h\}$ and 5 RBs are used for video data denoted by $\{\mathbf{a}_v,\mathbf{b}_v,\mathbf{c}_v,\mathbf{d}_v,\mathbf{e}_v\}$ in the layer of $L_h$, where the highest-SNR RBs in layer $L_h$ for haptic data are set to be $\{1,5,6\}$.  The permutation matrix may be then formed as follows:
\begin{align}
&\mathbf{\Pi}_{L_h} \mathbb{X} (L_h)   = \nonumber \\
&  \begin{pmatrix} \mathbf{I} & \mathbf{0} & \mathbf{0} & \mathbf{0} & \mathbf{0} & \mathbf{0} & \mathbf{0} & \mathbf{0}\\
\mathbf{0} & \mathbf{0} & \mathbf{0} & \mathbf{I} & \mathbf{0} & \mathbf{0} & \mathbf{0} & \mathbf{0}\\
\mathbf{0} & \mathbf{0} & \mathbf{0} & \mathbf{0} & \mathbf{I} & \mathbf{0} & \mathbf{0} & \mathbf{0}\\
\mathbf{0} & \mathbf{0} & \mathbf{0} & \mathbf{0} & \mathbf{0} & \mathbf{I} & \mathbf{0} & \mathbf{0}\\
\mathbf{0} & \mathbf{I} & \mathbf{0} & \mathbf{0} & \mathbf{0} & \mathbf{0} & \mathbf{0} & \mathbf{0}\\
\mathbf{0} & \mathbf{0} & \mathbf{I} & \mathbf{0} & \mathbf{0} & \mathbf{0} & \mathbf{0} & \mathbf{0}\\
\mathbf{0} & \mathbf{0} & \mathbf{0} & \mathbf{0} & \mathbf{0} & \mathbf{0} & \mathbf{I} & \mathbf{0}\\
\mathbf{0} & \mathbf{0} & \mathbf{0} & \mathbf{0} & \mathbf{0} & \mathbf{0} & \mathbf{0} & \mathbf{I}
\end{pmatrix} \begin{pmatrix} \mathbf{a}_h\\
\mathbf{b}_h\\
\mathbf{c}_h\\
\mathbf{a}_v\\
\mathbf{b}_v\\
\mathbf{c}_v\\
\mathbf{d}_v\\
\mathbf{e}_v
\end{pmatrix} = \begin{pmatrix} \mathbf{a}_h\\
\mathbf{a}_v\\
\mathbf{b}_v\\
\mathbf{c}_v\\
\mathbf{b}_h\\
\mathbf{c}_h\\
\mathbf{d}_v\\
\mathbf{e}_v
\end{pmatrix},
\end{align}
where $\mathbb{X}(L_h)$ is the un-permuted coded modulated symbols mapped to layer $L_h$.  The order of columns  4, 5, 6, 7 and 8 in $\mathbf{\Pi}_{L_h}$ may be arbitrarily changed without compromising performance as end-to-end mutual information remains unchanged.

\subsection*{\textbf{Step 4: Spatial Precoding}}\label{sec:Step4}
The spatial precoding is done in a conventional manner using Time-Division Duplex (TDD),  wherein the estimated uplink channel is used for downlink channel precoder computation.
For the channel at time instant $t$ in subcarrier $f$, $\mathbf{H}_{t,f}$, its Singular Value Decomposition (SVD) is given by
\begin{align}
\mathbf{H}_{t,f}=\mathbf{U}_{t,f} \mathbf{\Sigma}_{t,f}    \mathbf{V}^{\dagger}_{t,f} \in \mathbb{C}^{n_r \times n_t},
\end{align}
where the number of transmit and receive antennas are denoted as $n_t$  and $n_r$
{and $\dagger$ denotes conjugate transpose}. The matrices $\mathbf{U}_{t,f}$, $\mathbf{\Sigma}_{t,f}$ and   $\mathbf{V}_{t,f}$ denote the components of the SVD \cite{Tse05} of $\mathbf{H}_{t,f}$.  The $L$-layer precoder is
\begin{align}
\mathbf{P}_{t,f}=\tfrac{1}{\sqrt{L}}\mathbf{V}_{t,f} [1:L] \ \in \mathbb{C}^{n_t \times L},
\end{align}
where $\mathbf{V}_{t,f} [1:L]$ denotes the first $L$ vector of $\mathbf{V}_{t,f}$ corresponding to the highest eigenvalues.

Putting all steps together; the transmit signal prior to RE mapping and OFDM signal generation, can be then written:
\begin{align}
\mathbf{X}_{\texttt{t}}= \mathbf{P}\odot \mathbf{\Pi} \ \mathbb{X}(L_h,B_1),
\end{align}
where $\odot$ denotes element-wise product and the subscripts $t$ and $f$ are dropped in order to simplify the notation.
The RE mapping and  OFDM signal generation in Fig.~1 are done as those in NR \cite{TS38211}.


\section{Performance Evaluations}\label{sec:Peformance_evaluation}
In this section, we first present analytical capacity results for Rayleigh fading channels and then discuss  Monte-Carlo simulations under 3GPP channel models.

\subsection{Analytical Capacity Results}\label{sec:Peformance_Analysis}

%
Consider a BaseStation (BS) equipped with a large number of antennas (i.e. $n_t\gg 1$).
The channel matrix $\mathbf{H}$ of size ${n_r\times n_t}$ has spatial correlation matrix \mbox{$\mathbf{R}_{\mathbf{H}}= \mathbb{E} [\text{vec}(\mathbf{H}) \ (\text{vec}(\mathbf{H}))^{\dagger}]=\mathbf{R}_t^{\dagger} \otimes \mathbf{R}_r$}, where $\mathbf{R}_t$ and $\mathbf{R}_r$ are the TX/RX spatial correlation matrices, $\otimes$ and  $\text{vec}(\cdot)$ denote  Kronecker product and vectorization.
Then, the  eigenvalues of $\mathbf{H}$, $\pmb{\lambda}=[\lambda_1, \lambda_2, \cdots ]$, with exponential correlations at the arrays, follow a Gaussian distribution described as \cite{MIMO_correlated}
\begin{align}
\sqrt{n_t}\left({\pmb{\lambda}} - \bar{\pmb{\lambda}} \right)\sim \Ncal \left( \bar{\pmb{\lambda}} , \mathbf{C} \right),
\end{align}
where $\bar{\pmb{\lambda}}=[\bar{\lambda}_1, \bar{\lambda}_2, \cdots ]$ denotes their mean values. $\mathbf{C}$ is a {diagonal eigenvalue covariance matrix with} non-zero elements
\begin{align}
 (\mathbf{C})_{ii} =  {\lambda}_{r,i}^2 \|\Qbf  \Rbf_t \|_F^2 n_t,
\end{align}
where $\Qbf$ is the covariance matrix for the input symbols and ${\lambda}_{r,i}$ is the $i$th eigenvalue of $\Rbf_r$.

{We next derive the  instantaneous achievable rates for NR and MMCT. We consider a particular implementation of the scheme in Fig.~1 with two spatial layers (i.e.,  $L=2$) and given eigenvalues $\lambda_1, \lambda_2$.
After SVD, under complete channel state information at both the receiver and the transmitter for a MIMO setup} with $n_t\gg1$, the sum of instantaneous achievable rates for the two spatial layers can be written as follows:
\begin{align}
 r_{\text{NR}} =  \log \left( 1 + {\lambda}_1  \right) + \log \left( 1 + {\lambda}_2  \right).
\end{align}
{With MMCT, the} high-SNR layer is shared between the haptic and video streams (i.e. $L_h=1$).  In that layer, $B_1$ out of $B$ RBs are assigned to the haptic stream. Cf. \eqref{MMCT_Map} for an illustration of the space-frequency mapping. By assuming that $n_t\gg 1$, we can respectively obtain the achievable  rates for haptic and video streams as {follows:}
\begin{align}
 r_{\text{MMCT,h}} &=  \tfrac{B_1}{B} \log \left( 1 + {\lambda}_1  \right) {,}\\
 r_{\text{MMCT,v}} & =  \tfrac{B-B_1}{B} \log \left( 1 + {\lambda}_1  \right) +   \log \left( 1 + {\lambda}_2  \right),
\end{align}
where  ${\lambda}_1 \geq {\lambda}_2$.

\newcounter{mytempeqncnt2}
\begin{figure*}[t]
\normalsize
\setcounter{equation}{22}
\vspace{-.3cm}
\begin{align}\label{eq:out_pro_ana_NR}
    \mathbb{P}^{\texttt{out} }_\texttt{NR}=
    \begin{cases}
      1, & \hspace{1cm} \text{if}\hspace{1.2cm} \snr < \frac{2^{\frac{1}{2}R_\texttt{NR}^{\texttt{target}}}-1 }{n_t} \\
       \dfrac{2}{\pi}\arccos \left(\dfrac{1}{n_t\snr} \bigg((1+n_t\snr)^2 -2^{R_\texttt{NR}^{\texttt{target}}}\bigg)^{\frac{1}{2}} \right), &  \hspace{1cm} \text{if}\hspace{0.2cm}\ \frac{2^{\frac{1}{2}R_\texttt{NR}^{\texttt{target}}}-1 }{n_t} \leq  \snr \leq  \frac{2^{R_\texttt{NR}^{\texttt{target}}}-1 }{2n_t}\\
      0, & \hspace{1cm} \text{if}\hspace{1.2cm} \snr > \frac{2^{R_\texttt{NR}^{\texttt{target}}}-1 }{2n_t}
    \end{cases}.
\end{align}
\vspace{-.7cm}
\hrulefill
\vspace*{-.5pt}
\end{figure*}
\setcounter{equation}{14}

{In order to derive the NR and MMCT capacities, we} further assume {uncorrelated transmission} (i.e. $\Rbf_t=\Ibf$) with a uniform power allocation  (i.e. $\Qbf=\Ibf P$, where $P$ being average transmit power per antenna) at the BS, and a correlated {received signal} with the receive correlation matrix
\begin{align}
& \Rbf_r (\theta)=  \begin{pmatrix} 1 & \cos(\theta) \\
\cos(\theta) & 1
\end{pmatrix},
\end{align}
where $-\tfrac{\pi}{2}< \theta \leq  +\tfrac{\pi}{2}$ models the angle that impacts the correlation at the receiver side. In particular, one may interpret $\theta$ as the azimuth angles of the arrival beams in a proper coordinate system. 
The eigenvalues of $\Rbf_r$ are then equal to $1 \pm \cos(\theta)$. For a given $\theta$, we use Jensen's inequality for the function $\log(1+x)$, to move the expectation inside the $\log(\cdot)$ argument to simplify the capacity formula.
This yields
\begin{align}\label{eq:NR_capacity1}
 C_{\text{NR}}(\theta)= \mathbb{E} [r_{\text{NR}}] &\lessapprox \log \Big( 1 + \big(1+\cos(\theta)\big)n_t \snr \Big) \nonumber \\
 &+ \log \Big( 1 + \big(1-\cos(\theta)\big)n_t \snr \Big).
\end{align}
As a sanity check, the cases with $\theta=\tfrac{\pi}{2}, 0$ respectively correspond to independent and fully correlated channels, whose capacities simplify to the following known expressions:
\begin{align}
& C_{\text{NR}}(\theta=\tfrac{\pi}{2})\lessapprox 2 \log \Big( 1 + n_t \snr \Big),   \\
& C_{\text{NR}}(\theta=0)\lessapprox \log \Big( 1 + 2 n_t \snr \Big).
\end{align}
The former one has a multiplexing gain acting as two parallel MISO channels (cf. Eq. (38) in \cite[Sec. V]{MISO_majid}) and the latter acts as a single MISO channel but with double power.
For the MMCT, we can similarly obtain
\begin{align}
 C_{\text{MMCT,h}}(\theta) &= \mathbb{E} [r_{\text{MMCT,h}}] \lessapprox  \tfrac{B_1}{B} \log \Big( 1 + \big(1+\cos(\theta)\big)n_t \snr \Big),\\
 C_{\text{MMCT,v}}(\theta)&=\mathbb{E} [r_{\text{MMCT,v}}]  \lessapprox \!\tfrac{B-B_1}{B}\! \log \! \Big(\! 1 + \big(1+\cos(\theta)\big)n_t \snr \!\Big) \nonumber \\
 & + \log \Big( 1 + \big(1-\cos(\theta)\big)n_t \snr \Big). \label{eq:MMCT_capacity2}
\end{align}

We next compute  the outage probability using the above capacity results, to assess the performance, defined as
\begin{align}\label{eq:out_pro}
\mathbb{P}^{\texttt{out} }_\texttt{scheme}=\text{Pr}\left(C_\texttt{scheme}(\theta) <R_\texttt{scheme}^{\texttt{target}}\right),
\end{align}
where $C_\texttt{scheme}$ is the capacity for a  given scheme and $R_\texttt{scheme}^{\texttt{target}}$ denotes a fixed transmission rate target for that scheme. Three cases are obtained  by setting $\texttt{scheme}\in \{\text{NR}$, $  \text{MMCT,h} \ , \  \text{MMCT,v} \}$, with capacities  in \eqref{eq:NR_capacity1}, (19), and \eqref{eq:MMCT_capacity2}.
For uniform distribution of angles, the outage probabilities can be numerically computed using
\begin{align}\label{eq:out_pro}
\mathbb{P}^{\texttt{out} }_\texttt{scheme}\approx \frac{1}{M} \! \sum\nolimits_{m=1}^{M}  \mathds{1} \! \Big[C_\texttt{scheme}\left( \frac{m\pi}{M} - \frac{\pi}{2}\right) <R_\texttt{scheme}^{\texttt{target}}\Big],  \hspace{.2cm}
\end{align}
for $M\gg 1$, where $\mathds{1}\![\cdot]$ denotes the indicator function which returns $1$ if the argument is true, otherwise it returns zero. It is also possible to obtain closed-form expressions using a similar methodology as that in \cite{khormuji_relay}. We here report the closed-from analytical expressions for NR and haptic stream in MMCT under a uniform distribution of the angles. The outage probability for the former is given in \eqref{eq:out_pro_ana_NR} on the top of this page, whose proof is omitted for brevity. We however present the analytical derivation  of the outage probability for  the haptic stream in the MMCT  to illustrate the method. Consider
\setcounter{equation}{23}
\begin{align}\label{eq:Pout_intermediate}
\mathbb{P}^{\texttt{out} }_\texttt{MMCT,h}&=\text{Pr}\left(C_\texttt{MMCT,h}(\theta) <R_\texttt{MMCT,h}^{\texttt{target}}\right)\nonumber\\
&=\text{Pr}\left(\frac{B_1}{B} \log \Big( 1 + \big(1+\cos(\theta)\big)n_t \snr \Big) <R_\texttt{MMCT,h}^{\texttt{target}}\right) \nonumber\\
&=\text{Pr}\left( \cos(\theta)  <\frac{2^{\frac{B}{B_1}R_\texttt{MMCT,h}^{\texttt{target}}}-1}{n_t \snr}-1\right).
\end{align}
We next calculate  the outage probability in \eqref{eq:Pout_intermediate} by examining  three distinct  cases:

\textbf{Case 1}: If the condition
\begin{align}
\frac{2^{\frac{B}{B_1}R_\texttt{MMCT,h}^{\texttt{target}}}-1}{n_t \snr}-1>1,
\end{align}
is met, the haptic stream undergoes a full outage with probability one since $\cos(\theta)\leq 1$. Thus,
\begin{align}
\text{if} \ \  \snr < \underline{\snr}_\texttt{h}:= \frac{2^{\frac{B}{B_1}R_\texttt{MMCT,h}^{\texttt{target}}}-1}{2 n_t} \ \  \Longrightarrow  \ \ P^{\texttt{out} }_\texttt{MMCT,h}=1.
\end{align}

\textbf{Case 2}:
If the condition
\begin{align}
\frac{2^{\frac{B}{B_1}R_\texttt{MMCT,h}^{\texttt{target}}}-1}{n_t \snr}-1<0,
\end{align}
is met, the haptic stream prevents the outage with probability one since $\cos(\theta)\geq 0$ for $-\tfrac{\pi}{2} < \theta\leq +\tfrac{\pi}{2}$. Thus,
\begin{align}
\text{if} \ \  \snr > \overline{\snr}_\texttt{h}:= \frac{2^{\frac{B}{B_1}R_\texttt{MMCT,h}^{\texttt{target}}}-1}{ n_t} \ \  \Longrightarrow  \ \ P^{\texttt{out} }_\texttt{MMCT,h}=0.
\end{align}

\textbf{Case 3}:
If the condition
\begin{align}
0\leq \frac{2^{\frac{B}{B_1}R_\texttt{MMCT,h}^{\texttt{target}}}-1}{n_t \snr}-1\leq 1,
\end{align}
is met, then
\begin{align}
\mathbb{P}^{\texttt{out} }_\texttt{MMCT,h}
&=\text{Pr}\left( \cos(\theta)  <\frac{2^{\frac{B}{B_1}R_\texttt{MMCT,h}^{\texttt{target}}}-1}{n_t \snr}-1\right)\nonumber\\
&=\text{Pr}\left( \arccos \left(\frac{2^{\frac{B}{B_1}R_\texttt{MMCT,h}^{\texttt{target}}}-1}{n_t \snr}-1\right) \leq \theta  \leq \tfrac{\pi}{2}\right) \nonumber\\
&+\text{Pr}\left(\!-\tfrac{\pi}{2}\! <  \! \theta  \! \leq \! \arccos\! \left(\tfrac{2^{\frac{B}{B_1}R_\texttt{MMCT,h}^{\texttt{target}}}-1}{n_t \snr}-1\right) - \tfrac{\pi}{2} \right)\nonumber\\
&=2 \cdot\text{Pr}\left( \arccos \left(\frac{2^{\frac{B}{B_1}R_\texttt{MMCT,h}^{\texttt{target}}}-1}{n_t \snr}-1\right) \leq \theta  \leq \tfrac{\pi}{2}\right)\nonumber\\
&=2\int_{\arccos \left(\frac{2^{\frac{B}{B_1}R_\texttt{MMCT,h}^{\texttt{target}}}-1}{n_t \snr}-1\right)}^{\frac{\pi}{2} } \frac{1}{\pi} d\theta  \nonumber\\
&=1- \frac{2}{\pi}\arccos \left(\frac{2^{\frac{B}{B_1}R_\texttt{MMCT,h}^{\texttt{target}}}-1}{n_t \snr}-1\right).
\end{align}

By combining the above three cases and using one last trigonometric identity for Case 3,  we thus obtain
\begin{align}\label{eq:out_pro_ana}
    &\mathbb{P}^{\texttt{out} }_\texttt{MMCT,h}\nonumber\\
    &=
    \begin{cases}
      1, & \text{if}\ \snr < \underline{\snr}_\texttt{h} \\
      \dfrac{2}{\pi}\arcsin \! \left(\!\tfrac{2^{\frac{B}{B_1}R_\texttt{MMCT,h}^{\texttt{target}}}-1}{n_t \snr}-1\!\right), &  \text{if}\ \underline{\snr}_\texttt{h} \! \leq \! \snr\! \leq\!  \overline{\snr}_\texttt{h}\\
      0, & \text{if}\ \snr > \overline{\snr}_\texttt{h}
    \end{cases}.
\end{align}

\begin{figure} \label{fig:outage}
\centering
\vspace{-0.2cm}
            \includegraphics[width=0.384\textwidth]{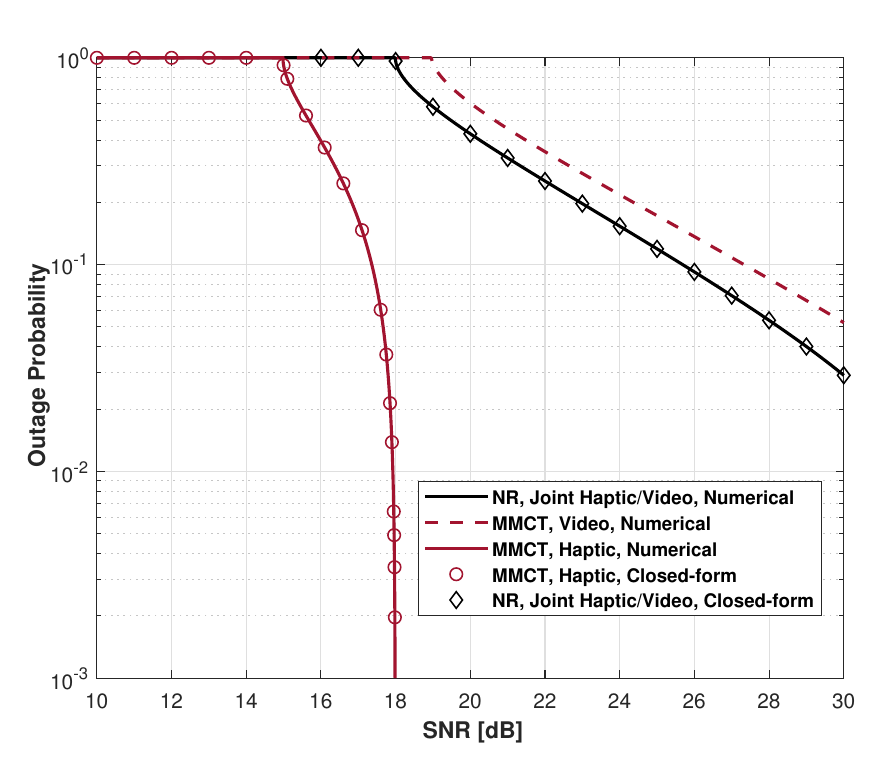}
            \vspace{-0.2cm}
                 \caption{Outage probabilities  of  NR  and  haptic and video in the proposed MMCT  with $L=2$, $L_h=1$, and $B_1=0.1 B$.
                 }
     \vspace{-0.5cm}
\end{figure}

Fig.~2 depicts  the outage probability as a function of SNR normalized by the number of transmit antennas when aiming for a total target transmission rate of 12 bits/s/Hz for two layers, (i.e. $R_\texttt{NR}^{\texttt{target}}=12$) for a scenario with  a uniform distribution of angles $\theta$. We set $B_1=0.1B$  and to make a fair comparison we set $R_\texttt{MMCT,h}^{\texttt{target}}=(12/L)\cdot\tfrac{B_1}{B}$ since one layer is used for the haptic data. We set $R_\texttt{MMCT,v}^{\texttt{target}}=12$ as that in the NR baseline.  The plots are \emph{numerically} obtained using \eqref{eq:out_pro}. For the haptic data and NR, the numerical evaluations  perfectly match the \emph{closed-form} analytical  expressions in \eqref{eq:out_pro_ana} and \eqref{eq:out_pro_ana_NR}, respectively.  For the haptic stream, the minimum normalized SNR (i.e. $n_t\overline{\snr}_\texttt{h}$) to avoid an outage is nearly $18$ dB, while it is above 30 dB for the NR joint transmission.  We observe that the proposed MMCT solution is able to produce a haptic stream with much lower outage probability, resulting in a more reliable haptic stream. This substantial enhancement is harvested  at a cost of a slight increase in the outage probability  for the video stream compared to the NR joint transmission. However, this increase is not a significant hindrance to system performance, especially considering the higher BLER demanded by the video stream. Thus, the described MMCT scheme can \emph{concurrently} generate  two streams, one of which  having a \emph{notably lower} outage probability, suitable  for haptic data, ultimately enabling a robust multi-modal transmission. The key enabler for such a desirable performance is the correlated array at the receiver, cf.  (15). In other words, we cannot achieve an improved outage probability for haptic transmission if the reception array at the UE  is uncorrelated, which can be seen by comparing (16)  and (19) for fixed $\theta=\frac{\pi}{2}$.

\subsection{Monte-Carlo Simulations}\label{sec:Peformance_simulation}

Table~I lists the main simulation parameters.
We consider a total frequency allocation with $B=20$ RBs with subcarrier spacing 30 kHz (which corresponds to a 7.2 MHz signal bandwidth).
One OFDM symbol in each slot is used for transmission of haptic and video information.
The haptic information is mapped to a subset of RBs having higher SNR compared to other
RBs wherein a data stream requiring low BLER is transmitted.
There are $L=2$ spatial layers and the haptic {stream} uses only one layer {--} $L_h=1$ in MMCT .
The carrier frequency is  $f_c=3.5$ GHz.
The channel model is CDL-C with $100$ ns delay spread.
The UE speed is set to $1$ m/s which is a typical speed for tactile internet evaluations.
The transmit antenna array at the gNB is assumed to be a panel consisting of $4\times 4$ uniformly spaced cross-polarized antenna elements with half-wavelength inter-element distance (a typical NR configuration). Each antenna element corresponds to one antenna port.
The receive antenna array at the UE is assumed to be a panel consisting of $2\times 1$  cross-polarized antenna elements with half-wavelength spacing.
{There is no coupling between the antenna elements.}
In the MMCT,  we transmit the same total number of information bits and use
the same total amount of resources as NR for both haptic and video data,  obtaining the same total spectral efficiency.
{Realistic channel traces have been produced according to
one of the most challenging multipath propagation scenarios specified by 3GPP.
Evaluations using real-world channel traces are beyond the scope of this work.}

\begin{table}[t!]\label{Table_Simulation}
  \centering
  \vspace{-.0cm}
  \caption{Simulation Parameters}
  \vspace{-.2cm}
\begin{tabular}{| c | c |}
\hline \hline
Carrier Frequency & 	$f_c=3.5$ GHz \\
\hline	
Subcarrier Spacing &	30 kHz \\
\hline	
Bandwidth &	7.2 MHz (corresp. to 20 RBs) \\
\hline	
Rank 	& 2 \\
\hline	
Modulation &	256QAM\\
\hline	
Code Rate	&determined based on selected MCS\\
\hline	
Precoding &	SVD \\
\hline	
MIMO Rx	& Zero Forcing (ZF)\\
\hline	
Channel Model	&CDL-C\\
\hline	
Delay Spread &	100 ns\\
\hline	
UE speed	& 1 m/s\\
\hline	
$L_h$ & 	1\\
\hline	
$B_1$ &	4 RBs\\
\hline	
CSI delay &	5 ms\\
\hline	\hline	
\end{tabular}
\vspace{-.5cm}
\end{table}

Fig.~3 plots the performance of the following schemes:
\begin{itemize}
  \item \emph{\textbf{NR with haptic transmission alone}}

This  baseline allocates nearly 10\% of the bandwidth  to the haptic data, i.e. the bandwidth is divided proportionally to the ratio
between the haptic and the video data.

  \item  	\emph{\textbf{NR with  video transmission alone}}

This scheme represents the baseline when nearly 90\% of the bandwidth is allocated to the video data.

  \item \emph{\textbf{NR with joint haptic and video transmission}}

This baseline represents the case when mixture of haptic and video bits are jointly encoded, modulated and precoded using the total available bandwidth.

  \item 	\emph{\textbf{NR with haptic transmission alone at much lower MCS}}

This scheme is similar to the first one where 10\% of the bandwidth is allocated to the haptic data but the MCS is intentionally reduced to meet the BLER of 0.1\%.

  \item  	 \emph{\textbf{Haptic transmission using MMCT}}

This plot indicates the performance of the haptic data transmitted via the  MMCT scheme when the haptic data only occupies 10\% of time-frequency-space resources.

\item \emph{\textbf{Video transmission  using MMCT}}

This plot indicates the performance of the video data transmitted using the MMCT solution when the video data occupies the remaining 90\% of the  resources.

\end{itemize}

\begin{figure} \label{fig:performance1}
\centering
            \includegraphics[width=0.4\textwidth]{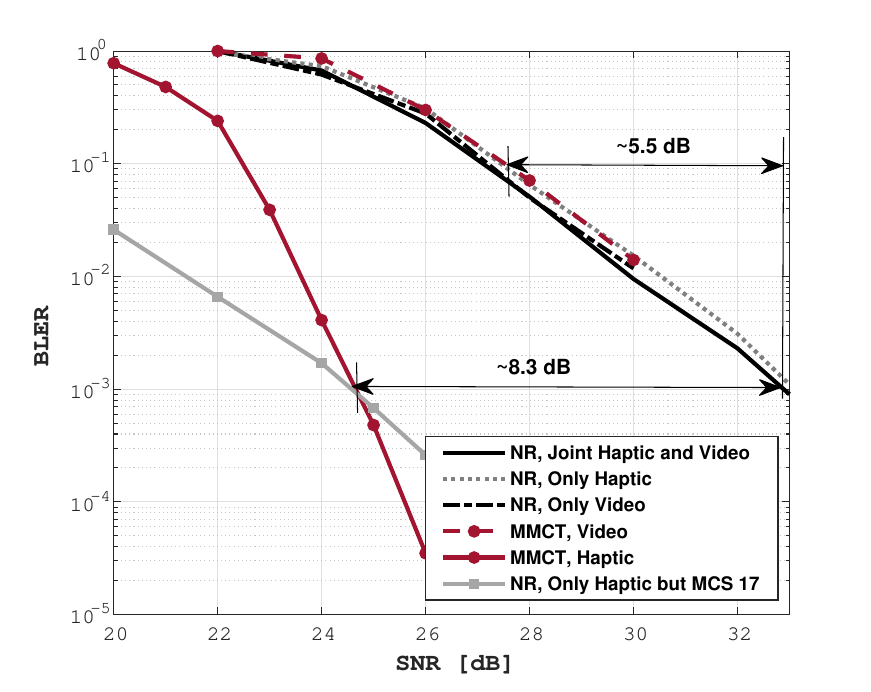}
            \vspace{-0.3cm}
                 \caption{ BLER performances of NR  baseline schemes and  the proposed MMCT solutions with $L=2$ and $L_h=1$.}
     \vspace{-0.6cm}
\end{figure}


This evaluated case  represents a very high data rate transmission wherein two spatial layers are used.
The MCS index is set to 25 which corresponds to $13.82$ bits/s/Hz for the two-layer transmission-- see \cite{TS38214}.
In this evaluation, the haptic data is 10\% of the total data. The layer mapper for this case, is configured according to $B_1=4,L_h=1$.


The power gain of the MMCT solution can be obtained using the results in Fig.~3. We first need to find the lowest SNR for which both haptic and video data can meet their requirements of $10\%$ and $0.1\%$. In this case we see that, if we set the SNR close to $27.4$ dB, we meet both requirements. The best NR solution with the same MCS happens to be the joint transmission at SNR of $32.9$ dB, so as to meet the most stringent data requirement which is haptic with BLER of $0.1\%$. This results in a power gain $\mathbf{Gain}_{\texttt{eff}}=  5.5$ dB  as marked in the figure. For the haptic data we have a higher gain $\mathbf{Gain}_h=  8.3$ dB as marked in the figure. The grey solid line corresponds to haptic transmission with the highest MCS -- MCS 17 -- that provides BLER $\leq 0.1\%$ at same SNR as haptic transmission within MMCT (solid red line). Transmission of haptic data with MCS 17 corresponds to a $35\%$ spectral efficiency loss compared to haptic transmission within MMCT.

Fig.~4 shows the corresponding BER of the aforementioned schemes. The haptic transmission using the MMCT  offers an 8 dB power gain for the BER of $10^{-4}$ compared to that of the NR  with joint transmission. We also observe  that the video transmission of the new solution almost has \emph{no} loss compared to NR. As for the latency of the scheme, the enabled lower first-transmission BLER for the more demanding stream yields a reduced average number of retransmissions and ultimately results in a shorter latency. More accurate evaluations will be carried out in future investigations.


\section{Concluding Remarks}\label{sec:concl}
We presented a new physical-layer method for efficient transmission of multiple data streams with different reliability-latency performance requirements. The data from multiple streams are arranged within a same physical-layer transport block and jointly mapped to MIMO spatial layers and frequency resources according to the SNR of each resource so as to obtain the required performance boost for the more demanding streams.
Our findings illustrate that, when applied to a tactile internet application, the MMCT method delivers  significant SNR and spectral efficiency improvements  when compared to conventional 3GPP NR transmission methods.


\begin{figure} \label{fig:performance2}
\centering
            \includegraphics[width=0.4\textwidth]{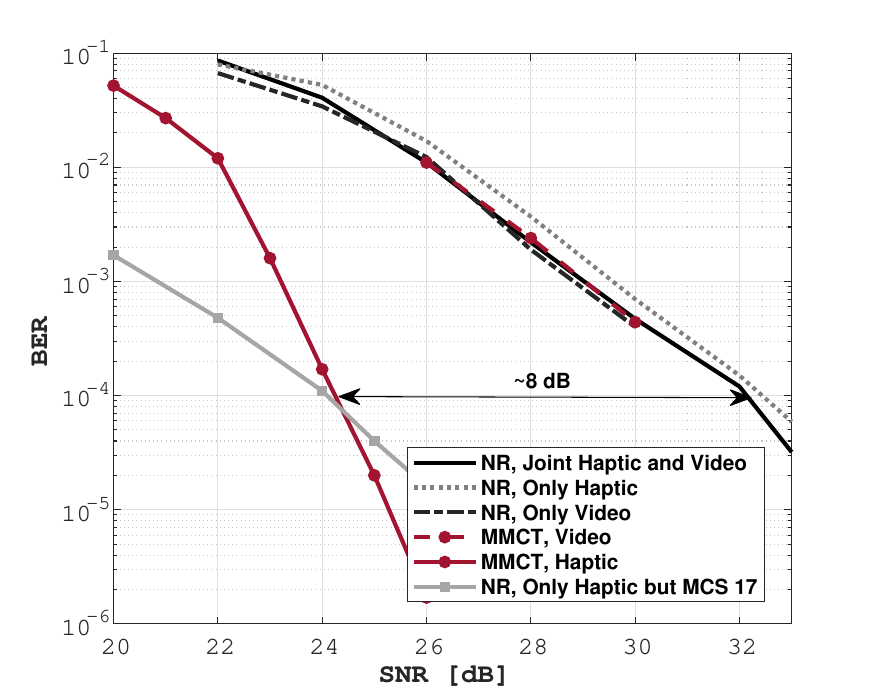}
              \vspace{-0.3cm}
                 \caption{ BER performances for the setups in Fig.~3. }
     \vspace{-0.5cm}
\end{figure}


\end{document}